\documentclass[preprint,showpacs,superscriptaddress,groupedaddress]{aastex61}  

\usepackage{graphicx}
\usepackage{dcolumn}
\usepackage{bm}
\usepackage{amssymb}
\usepackage{enumitem}
\usepackage{tikz}

\begin{document}

\title{Deep searches for broadband extended gravitational-wave emission bursts by heterogeneous computing}

\author{Maurice H.P.M. van Putten\footnote{Corresponding Author's Email: mvp@sejong.ac.kr}}
\affil{Yeongsil-Gwan, Room 614, Physics and Astronomy, Sejong University, Seoul, South Korea}


\begin{abstract}
We present a heterogeneous search algorithm for broadband extended gravitational-wave emission (BEGE), expected from gamma-ray bursts and energetic core-collapse supernovae. It searches the $(f,\dot{f})$-plane for long duration bursts by inner engines slowly exhausting their energy reservoir by matched filtering on a {\em Graphics Processor Unit} (GPU) over a template bank of millions of one-second duration chirps. Parseval's Theorem is used to predict the standard deviation $\sigma$ of filter output, taking advantage of near-Gaussian noise in LIGO S6 data over 350-2000 Hz. Tails exceeding a mulitple of $\sigma$ are communicated back to a {\em Central Processing Unit} (CPU). This algorithm attains about 65\% efficiency overall, normalized to the Fast Fourier Transform (FFT). At about one million correlations per second over data segments of 16 s duration $(N=2^{16}$ samples), better than real-time analysis is achieved on a cluster of about a dozen GPUs. We demonstrate its application to the capture of high frequency hardware LIGO injections. This algorithm serves as a starting point for deep all-sky searches in both archive data and real-time analysis in current observational runs.
\end{abstract}

\nopagebreak


\section{Introduction}

Gravitational radiation offers a potentially powerful new channel to discover the physical nature and population statistics of core-collapse supernovae and their association with neutron stars and black holes. Recently, LIGO identified a black hole binary progenitor of GW150914 \citep{lig16} with remarkably low spin. Stellar mass black holes are believed to be remnants of extreme transient events such as gamma-ray bursts and core-collapse supernovae. Shortly after birth in the latter, possibly including the superluminous variety, black holes may encounter strong interactions with high density matter. This outlook opens a window to release their angular momentum in gravitational radiation, leaving slowly spinning remnants with $a/M\simeq 0.3$ in dimensionless spin \citep{van17}. Future detection of similar events may reveal whether GW150915 is typical or the tail of a broad distribution in black hole mass and spin. 

Neutron stars and black holes born in core-collapse of massive stars are of great interest as candidate sources of gravitational waves, especially for their potential to be visible also in the electromagnetic spectrum. Searches for these events may be triggered in either radiation channel \citep{sat09,van16} and a combined detection would enable identification of source and host environment, in the footsteps of multi-wavelength observations of gamma-ray bursts pioneered by {\em BeppoSAX} \citep{fro09}. EM-triggers obtained from transient surveys allow off-line GW-analysis of LIGO-Virgo archive data. On the other hand, GW-triggers require relatively low latency in EM follow-up, that may be challenging by modest localisation of LIGO-Virgo detections.

While core-collapse supernovae are relatively numerous, only a small fraction is known to be associated with extreme events. For instance, the true event rate (corrected for beaming) of long GRBs is about 1 per year within a distance of 100 Mpc. Achieving sensitivity to tens of Mpc to emissions limited to $E_{GW}=O\left(1M_\odot c^2\right)$, where $c$ denotes the velocity of light, poses a challenge for {\em deep searches} in gravitational wave data. 

Broadband extended gravitational-wave emission (BEGE) from aforementioned extreme events may be produced with durations lasting up to tens of seconds. In chirp-based spectrograms, such may appear as trajectories marked by frequencies slowly wandering in time, featuring ascending and descending chirps \citep{lev15}. To search for these signatures in the $(f,\dot{f})$-plane, we recently devised a dedicated butterfly filtering using chirp templates of intermediate duration $\tau$ of, e.g., one second, targeting a time scale of phase coherence that may capture tens to hundreds of wave periods associated with non-axisymmetric accretion flows. Using millions of chirp templates it can detect complex signals such as Kolmogorov scaling in noisy time-series, recently in {\em BeppoSAX} gamma-ray light curves with an average photon count down to 1.26 photons per 0.5 ms bin \citep{van14a}. This kind of sensitivity suggests to explore its further applications to strain amplitude gravitational wave data \citep{van16}. 

Deep searches covering a complete science run of LIGO requires considerable computing resources in the application of butterfly filtering with a dense bank of templates. We here report on a novel algorithm by heterogeneous computing comprising both {\em Graphics} and {\em Central Processing Units} (GPUs, respectively, CPUs) using the {\em Open Compute Language} (OpenCL) \citep{gas11,khr15}.

A primary challenge in heterogeneous computing is circumventing GPU-CPU bottlenecks arising from potentially vast discrepancies in data throughput over the {\em Peripheral Component Interface} (PCI). Our algorithm exploits near-Gaussian noise in the high frequency bandwidth of 350-2000 Hz in LIGO data, whereby the output of matched filtering is essentially Gaussian as well. Near-optimal efficiency is obtained by retaining only tails of relatively high signal-to-noise ratios back to the CPU from the GPU output, whose cut-off is predicted by Parseval's Theorem. Including overhead in the latter, our algorithm achieves about 65\% efficiency normalized to GPU-accelerated Fast Fourier Transform (FFT) in {\em complex-to-complex} (C2C), {\em single precision} (SP) and {\em interleaved out-of-place} memory allocation.

Our choice of chirp templates is guided by inner engines involving black holes described by the Kerr metric, interacting with high density matter, expected in core-collapse of massive stars and mergers involving neutron stars, the latter envisioned in association with short GRBs with Extended Emission (SGRBEE) and long GRBs with no supernovae (LGRBN) such as GRB060614 \citep{van14b}. 

In the application to LIGO S6, we give a detailed description of our data-base, which comprises bandpass filtering to aforementioned 350-2000 Hz (over 64 s segments of data, $N=2^{18}$ samples) and a restriction to simultaneous H1-L1 detector output (29.4\% of total S6 data). 
On a modern GPU, we realize approximately 80,000 correlations per second over 16 s segments of data ($N=2^{16}$ samples). On a cluster of about a dozen GPUs, about 1 million correlations per second realizes better than real time analysis. As such, the presented method is applicable to both archive analysis and low-latency searches in essentially real-time, pioneered following GW150914 \citep{abb16A} and for current Advanced LIGO runs \citep[e.g.][]{guo17}. For the present archive analysis of LIGO S6, however, our focus is on deep searches for an exhaustive search, all-sky and blind without triggers from electromagnetic data. 

Existing comprehensive, blind searches for bursts \citep{abb04,abb07,abb09a,abb09b,aba10,aba12,and05,aas13} cover various broadband emissions over 16-500 Hz \citep{abb17}, 40-1000 Hz \citep{abb16B} and, for short bursts, 32-4096 Hz \citep{abb17b}. Around a rotating black hole of mass $M$ newly formed in core-collapse supernovae, gravitational wave emission from non-axisymmetric mass-flow around the Inner Most Stable Circular Orbit (ISCO) is expected to be potentially luminous \citep{van01}, featuring a broadband descending chirp \citep{van08,van16} with late-time frequency \citep{van12}
\begin{eqnarray}
f_{GW} \simeq (595-704)\,\mbox{Hz}\left( \frac{10 M_\odot}{M}\right),
\label{EQN_fGW}
\end{eqnarray}
where the range in the frequency refers to dependency on initial black hole spin.
This motivates our present focus on the high-frequency bandwidth 350-2000 Hz in LIGO S6. In this frequency bandwidth, LIGO noise is essentially Gaussian,
which shall be exploited in our GPU-CPU method of analysis.

In \S2 we review chirp-based spectrograms by butterfly filtering. In \S3, our heterogeneous computing algorithm is described with use of {\em pre- and post-callback} functions. Benchmark results are given in \S4. \S5 reports on a detection of some illustrative LIGO S6 hardware burst and calibration injections. We summarize our findings and outlook in \S6.

\section{Bandpass filtered H1 and L1 data in S6}

\begin{figure}
\center{\includegraphics[height=7.3cm,width=8.cm]{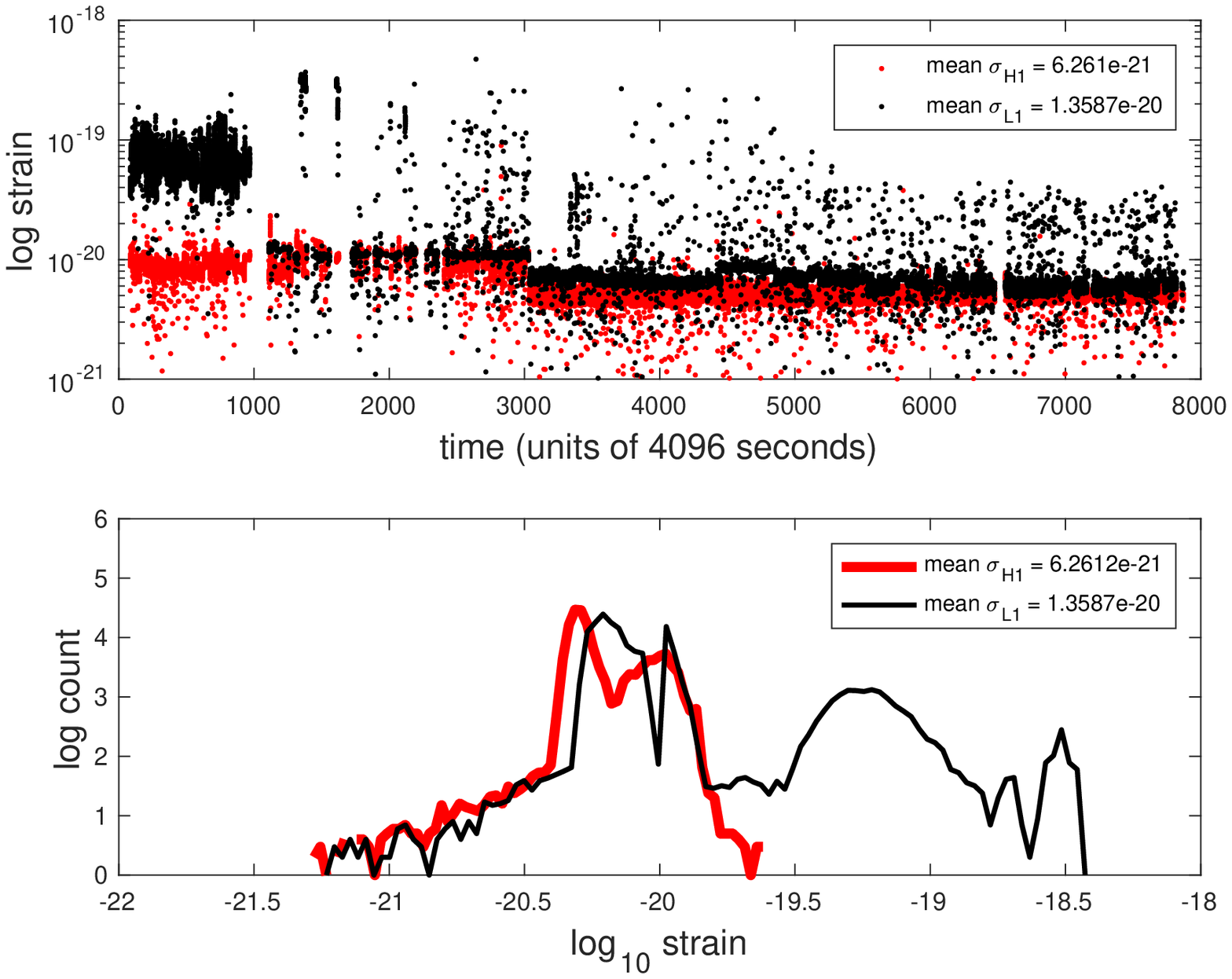}\hfill\includegraphics[height=7.3cm,width=8.cm]{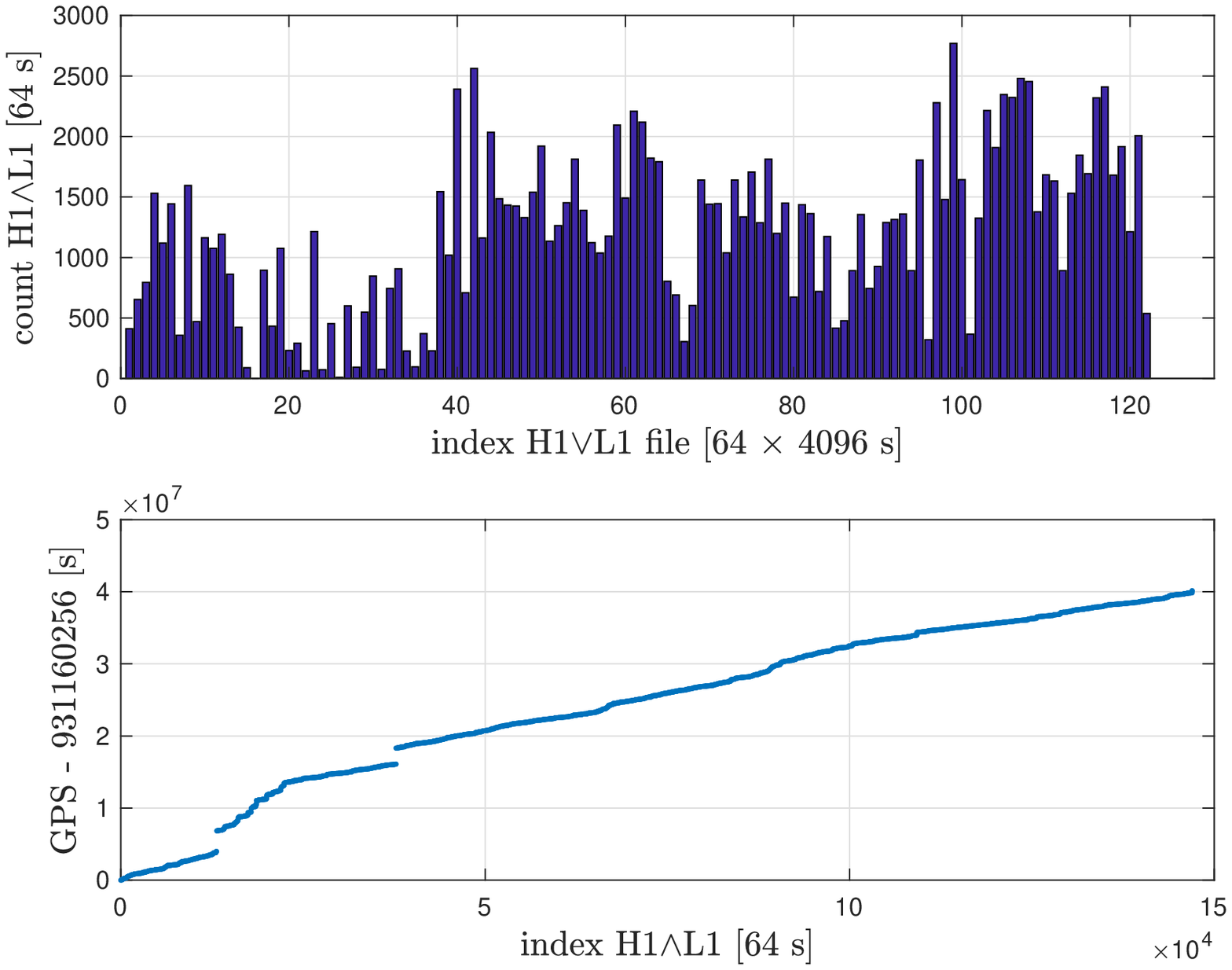}}
\caption{Overview of LIGO S6, showing standard deviations over 64 s data segments ($N=2^{18}$ samples) of H1$\land$L1. H1$\land$L1 ranged from 0-68\% with an average yield of 29.4\% of H1$\lor$L1. Strain noise  H1 and L1 was better than $10^{-20}$ over 89\%, respectively, 64\% of the time. Performance of H1 and L1 became somewhat more consistent after the first 3000 hours.}
\label{fig1}
\end{figure}

LIGO S6 covers the period July 7 2009 through October 20 2010. 
In our analysis of LIGO S6, we focus on epochs when H1 and L1 are both taking data.
These H1$\wedge$L1 data represent 29.4\% of data when either H1 or L1 were taking data
(H1$\lor$L1), measured over 64 second segments (Fig. \ref{fig1}).

In our search for gravitational wave emission from core-collapse supernovae associated with stellar mass black holes, we focus on the frequency bandwidth of 350-2000 Hz. Bandpass filtering (over 64 s segments of data, $N=2^{18}$ samples), LIGO noise is essentially Gaussian \citep[e.g.][]{van16}. This bandwidth may contain gravitational wave emission from non-axisymmetric mass motion about the Inner Most Stable Circular Orbit (ISCO) around stellar mass black holes \citep{van12,van16}. 

\begin{table}
{\bf Table 1.} {Overview of the data-base of H1$\land$L1 when both H1 and L1 were taking data (measured over 64 s data segments), 
extracted from a total of 12726 LIGO S6 frames. Frames on the LIGO Open Science Center (LOSC)
comprise 4096 s ($N=2^{24}$ samples) of H1 or L1 data, here bandpass filtered to 350-2000 Hz over 64 s data segments $(N=2^{18}$ samples). 
H1$\land$L1 data for analysis is in 36 files of $4096\times64$ s segments (Table 2).}
\center{{\begin{tabular}{@{}cccccc@{}}
\mbox{}\\\hline\hline
 	Data             & 64 s segments & LOSC frames (4096 s) & Memory & Source, Target\\
	\hline
	H1                		& 422912  & 6608 &  - & LOSC\\
	L1                 		&  391552 & 6118  & -  & LOSC\\
	H1$\lor$L1    		& 499712 & 7867  & 1.05 TB & Disk\\ \hline
	H1$\land$L1  		& 147000 & - & 305 GB & Disk\\ 
 	File       & $4096$ & - & 8.59\,GB & Compute node\\
	\hline
	\end{tabular}}}
	\label{Table1}
\end{table}

\section{Butterfly filtering by heterogeneous computing}\label{sec:2}

\begin{figure}[h]
\centerline{\includegraphics[scale=0.5]{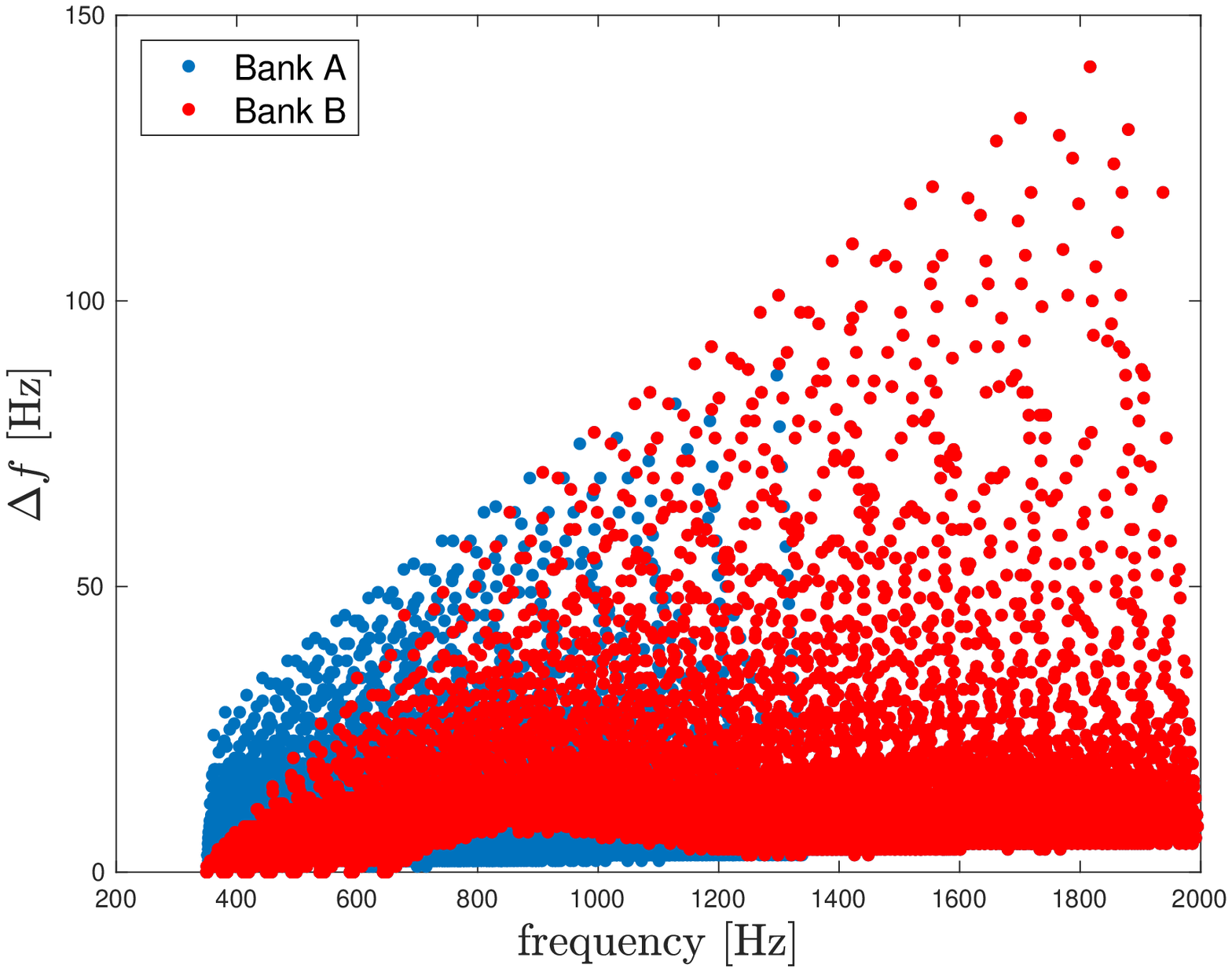}\hfill\includegraphics[scale=0.5]{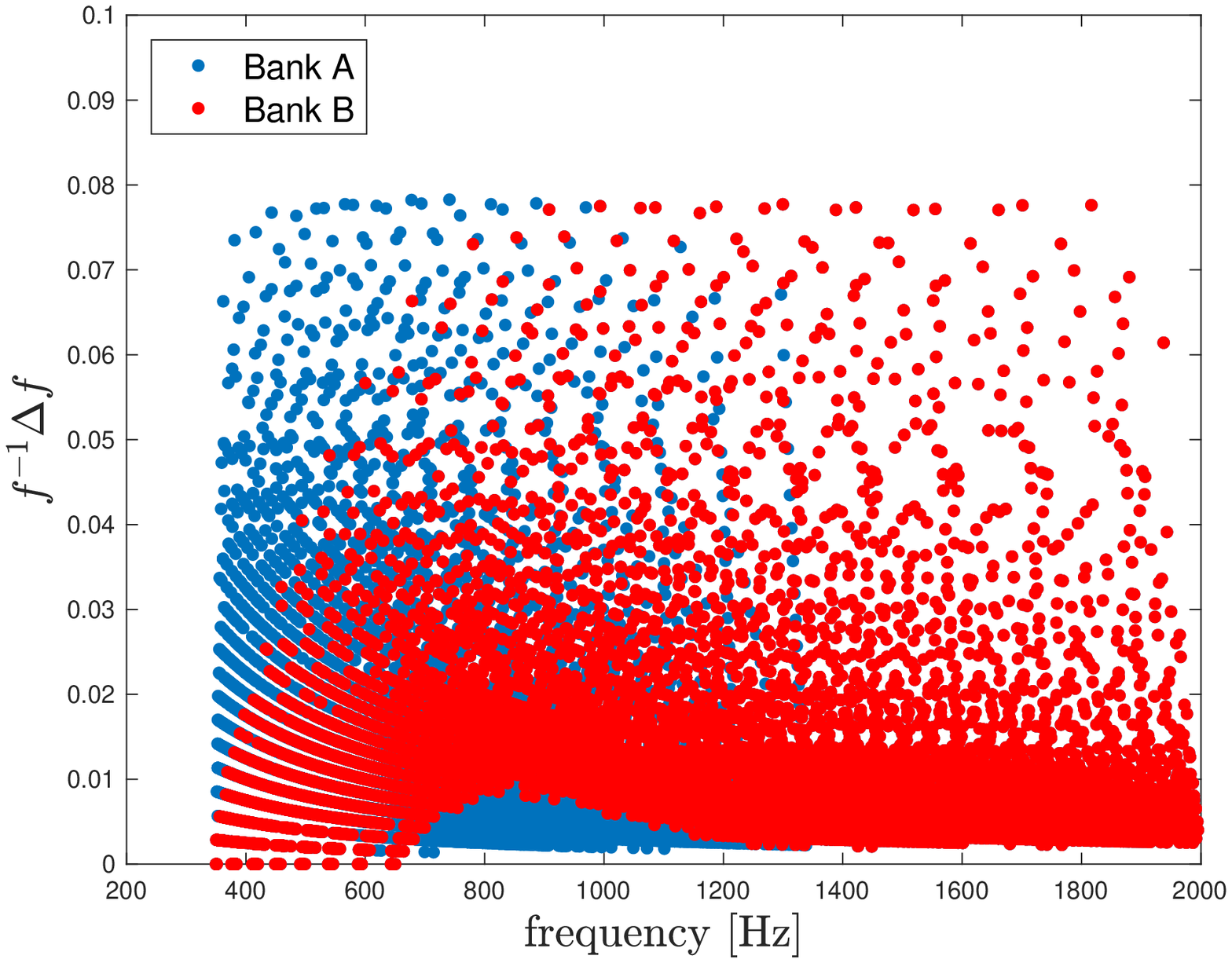}} 
\caption{Overview of template banks of one-second duration chirps (dots) covering 350-2000 Hz, shown by frequency $f$ and their change $\Delta f$ in frequency, illustrated with a bank of small size. The chirps used in butterfly filtering are symmetric in time, obtained by superposition of chirp forward and backward in time, suitable in searches for both ascending and descending chirps. Banks A and B are similar, except Bank A is larger by including more pronounced chirps (larger $\Delta f$) at the lower bound of 350 Hz.}
\label{figC}
\end{figure}

To search for slowly evolving trajectories in time-frequency space, we consider matched filtering over a large bank of chirp templates covering a range in $f$ and time rate-of-change of frequency $\dot{f}$, i.e., a butterfly
\begin{eqnarray}
0<\delta_1 < \left|\dot{f}\right|<\delta_2
\label{EQN_s}
\end{eqnarray}
for some $\delta_{1,2}>0$. Over a finite bandwidth of frequencies, the resulting output is a {\em chirp-based spectrogram}.
The chirps are generated from a long duration template, produced by solving a pair of ordinary differential equations modeling
black hole spin down against high density matter at the ISCO \citep{van14a}, the results of which are illustrated in Fig. \ref{figC}.

\begin{figure}[h]
\centerline{\includegraphics[scale=0.65]{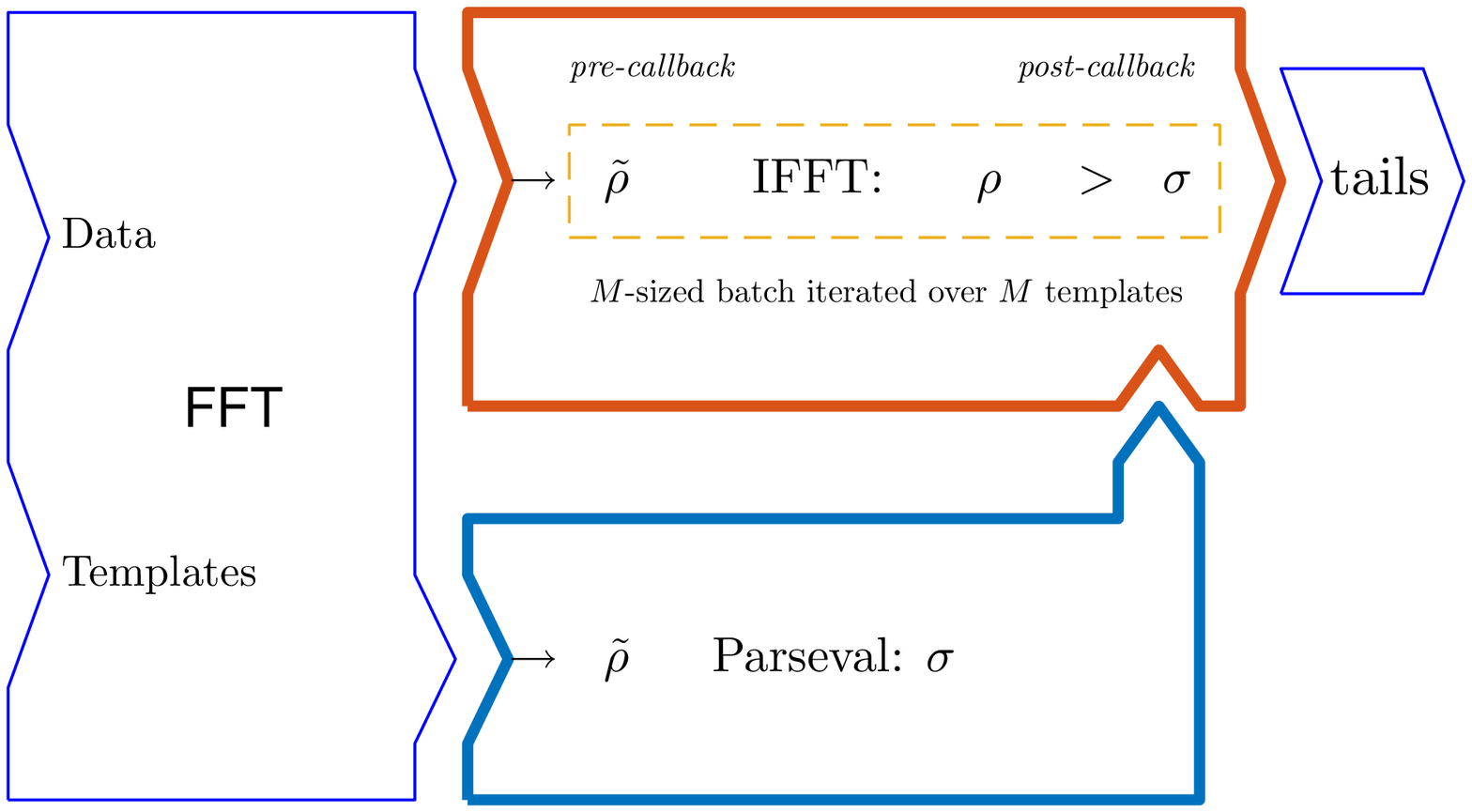}}
\hspace*{-0.4cm}\centerline{\includegraphics[scale=0.25,angle=-90]{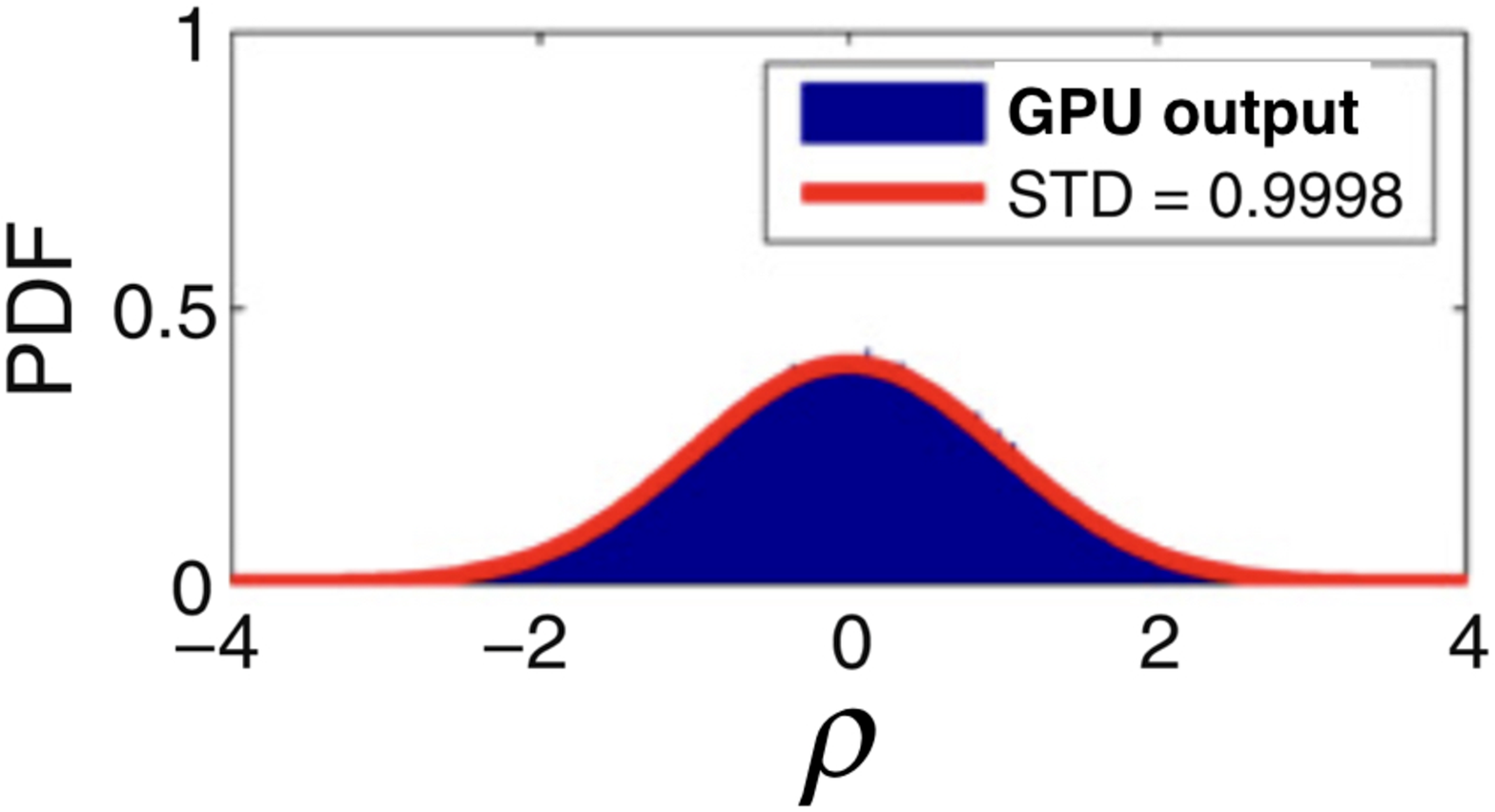}\includegraphics[scale=0.25,angle=-90]{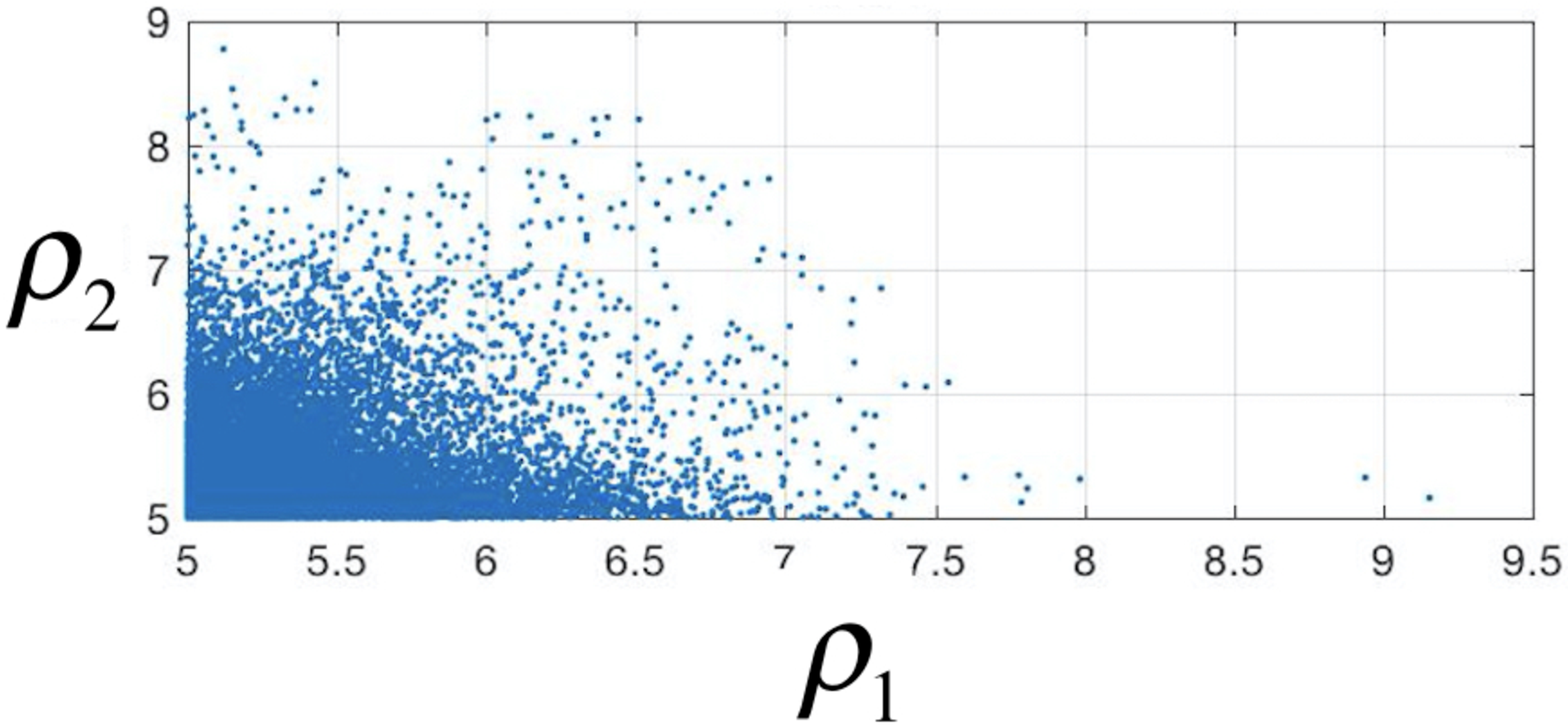}}
\caption{Butterfly filtering by heterogeneous computing applied to $M$ H1$\land$L1-data 16 s segments ($N=2^{16}$ samples) by CPU (thin lines) and GPU (thick lines). Parseval's Theorem computes $M^2$ standard deviations $\sigma$ of essentially Gaussian correlations $\rho(t)$ obtained by matched filtering on the GPU. Potentially relevant results are contained in the tails of $\rho(t)$. Retaining tails $\rho(t)>\kappa\sigma$ for a threshold $\kappa$ realizes near-optimal heterogeneous GPU-CPU computing, effectively circumventing PCI bandwidth limitations when $\kappa$ is a few.}
\label{fig3}
\end{figure}

Matched filtering of a time series $y(t)$ against chirps templates $w(t)$ is defined by correlations
\begin{equation}
\rho(t) = \int_{-\infty}^{\infty}{w(s)y(t+ s)dt}.
\label{EQN_A1}
\end{equation} 
In the present application to LIGO strain data, $y(t)$ and $w(t)$ have zero mean. This integral is conveniently
evaluated in the Fourier domain as $\tilde{\rho}(k)=\tilde{w}^*(k)\tilde{y}(k)$, where
\begin{eqnarray}
\tilde{f}(k)=\frac{1}{{2\pi}}\int_{-\infty}^\infty f(t) e^{-ikt}dt,~
f(t)=\int_{-\infty}^\infty \tilde{f}(k) e^{ikt}df.
\label{EQN_A2}
\end{eqnarray}
Discretizing (\ref{EQN_A1}) to samples at equidistant instances $t_n$ $(n=0,1,\cdots, N$), we evaluate (\ref{EQN_A2}) by FFT. This is more efficient compared to direct evaluation of (\ref{EQN_A1}) in the time domain, whenever the number of samples $N$ exceeds a few hundred. This may be readily observed by comparing compute times, convolving two vectors ${\bf u}$ and ${\bf v}$ by FFT versus direct evaluation in, e.g., {\em MatLab}; see also \cite{smi16}.

For reference, recall that correlating vectors ${\bf y}$ and ${\bf w}$ comprises three steps: twice forward FFT, pointwise products $\bf\tilde{\rho}$ involving complex conjugation, and one inverse FFT: 
\begin{eqnarray}
\{{\bf \tilde{w}},{\bf \tilde{y}}\}=\mbox{FFT}\{\bf w,\bf y\}, 
~\bf\tilde{\rho} = \bf\tilde{w}^*\cdot\bf\tilde{y},
~\bf\rho = \mbox{FFT}^{-1}\{\bf\tilde{\rho}\}.
\label{EQN_3S}
\end{eqnarray}
For LIGO S6, the (downsampled) sampling rate is 4096 s$^{-1}$, whence $N=2^{16}$ for 16 s data segments.

With vanishing mean values, the standard deviation of $\rho(t_n)$,
\begin{eqnarray}
\sigma = \frac{1}{\sqrt{N}}\sqrt{\sum_{n=0}^{N-1} \rho(t_n)^2},
\end{eqnarray}
satisfies Parseval's Theorem 
\begin{eqnarray}
\sigma = \frac{\sqrt{2}}{N}\sqrt{\sum_{n=1}^{N/2-1} \left| c_n\right|^2},
\label{EQN_PT}
\end{eqnarray}
where $c_n$ denote the Fourier coefficients of $\rho(t)$ according to the FFT pair
\begin{eqnarray}
c_k=\frac{1}{N}\sum_{n=0}^{N-1} f_n e^{-ikt_n},~~f_n=\sum_{k=0}^{N-1} c_k e^{ikt_n}.
\label{EQN_A3}
\end{eqnarray}

Bandpass filtered to 350-2000 Hz, H1$\land$L1 (Table 1) has noise which is essentially Gaussian \citep[e.g.][]{van16}. This property is inherited by $\rho(t)$ in (\ref{EQN_A1}). Hence, $\rho(t)$ is effectively described by $\sigma$ in (\ref{EQN_PT}) for a given pair of data segment and template. Therefore, (\ref{EQN_PT}) provides a  {\em predictive step} to the output of (\ref{EQN_3S}). In processing (\ref{EQN_3S}) on a GPU, a threshold in a {\em post-callback function} can be used to retain only tails (Fig. \ref{fig3})
\begin{eqnarray}
\rho(t_n) > \kappa \sigma
\label{EQN_tail}
\end{eqnarray}
for feedback to the CPU over the PCI. In (\ref{EQN_tail}), we implicitly apply the inequality to the absolute value of $\rho_n = \rho(t_n)$. Thus, (\ref{EQN_tail}) circumvents vast discrepancies in throughput of GPUs and CPUs whenever $\kappa$ is on the order of a few. This step is essential for an optimal heterogeneous computing algorithm, to be benchmarked further below.

It should be mentioned that below 350 Hz, LIGO data is non-Gaussian, giving rise to distributions of $\rho(t)$ that occasionally show multiple peaks. (This depends on the pair of data segment and template.) In this event, $\sigma$ inadequately describes $\rho(t_n)$, whereby tails defined by (\ref{EQN_tail}) become less meaningful in defining candidate detections. 

Processing is applied to batches of $M=2048$ of H1$\land$L1 16 s data. Such {\em block} of about 9 hours of data comprising about 1 GByte, suitable for allocation in {\em Global Memory} of a typical GPU. Chirp templates are extracted by time slicing from a model of black hole spindown \citep{van14a}. While these emissions are of relatively high frequency when the black hole spins rapidly, late time emission following spin down reaches an asymptotic frequency satisfying (\ref{EQN_fGW}). Analysis is performed in groups of $M$ such templates by FFT in {\em batch mode}. Batch mode operation is essential to reaching optimal FFT performance on a GPU. 

\begin{table}
	{\bf Table 2.} {Partitioning files of the H1$\land$L1 data-base on a heterogenous compute node into blocks allocated in Global Memory on a GPU for
	processing by FFT$_{N,M}$ with transforms of size $N=2^{16}$ in batch mode of size $M=2048$.}
	\center{	{\begin{tabular}{@{}llllcc@{}}
	\mbox{}\\\hline\hline
	Unit & Array length & Memory size & Target \\
	\hline
	File          & 8 blocks & 8.59\,GB & Disk storage, host\\ 
	Block      & $NM$ & 1.1\,GB & Global Memory/GPU  \\
	FFT batch size & $M=2048$ & 1.1 GB & FFT/GPU \\
	FFT data segment & $N=2^{16}$ & 0.5\,MB & Global and Local Memory/GPU \\
	\hline
	\end{tabular}}}
	\label{Table2}
\end{table}

\subsection{Teraflops compute requirements}

Sensitivity to arbitrary, slowly varying transients is realised by banks sufficiently large to densely cover the $(f,df/dt)$-parameter space. For matched filtering, a bank of chirps of one second duration covering $f = O(N)$\,Hz with frequency changes $O(f)$ will be dense with step sizes order of $1/N$\,Hz in $f$ and $df/dt$, setting a minimum bank size of order $O(N^2)$. For $f$ on the order of one kHz, the minimum bank size is $O(1\mbox{M})$, needed to ensure a reasonable probability to match a signal (a ``hit'' when $\rho > \kappa \sigma$).

For a better than real-time analysis by butterfly filtering of data segments of duration $T$ over a template bank of size $K_1M$, the required compute performance is
\begin{eqnarray}
\dot{n}=5N\log_2N\times K_1M T^{-1} = 2.75\,\mbox{teraflops},
\label{EQN_nd}
\end{eqnarray}
where the right hand side refers to our choice of $T=16$ seconds and a template bank of $\alpha=1,2,\cdots,K_1$ sets of size $M=2048$ each. 

Hardware requirements are considerably higher, since FFT's tend to be {\em memory limited} (not compute limited) on GPUs, especially when FFT array sizes exceed the size of {\em Local Memory} privy to individual {\em Compute Units} (CU). At typical efficiencies of $\eta\simeq 7\%$ in these cases, 
(\ref{EQN_nd}) points to a minimum requirement of about 50 teraflops at GPU maximal compute-performance, assuming 
(\ref{EQN_nd}) is realized at approximately optimal efficiency normalized to FFT.

In what follows, we consider partitioning template bank by $K_1M$ and, respectively, data in $\beta=1,2,\cdots,K_2$ blocks in
\begin{eqnarray}
W_{\alpha} = \{ w_{\alpha k}\}_{k=1}^M,~~Y_{\beta}= \{y_{\beta k}\}_{k=1}^M.
\end{eqnarray}
In our application, $K_1M=2^{23}$ (up to 8 million) $K_2M=288$ for LIGO S6. The total number of correlations for a full LIGO S6 analysis is
\begin{eqnarray}
K_1K_2M^2=5\times 10^{12}.
\label{EQN_K1K2}
\end{eqnarray}
For our choice of 16 second segments ($N=2^{16}$), (\ref{EQN_K1K2}) defines a compute requirement of $2.5\times 10^{19}$ floating point operations for a complete LIGO S6 analysis over a bank of 8M templates.

\subsection{Batch mode with pre- and post-callback functions}

Fig. \ref{fig3} shows the butterfly filtering by our GPU-CPU heterogeneous computing algorithm, based on detailed partitioning of data and work listed in Table 2. For (\ref{EQN_3S}), we choose FFT with C2C, SP and with interleaved out-of-place memory allocation by one-dimensional FFT$_{N,M}$ of length $N=2^{16}$ in batch mode of size $M=2048$:
\begin{enumerate}[label=(\roman*)]
\item FFT$_{N,M}$ of $M$ pairs of 16 s data segments of H1$\land$L1 comprising a block of $MN$ CSP in {Allocatable Memory} of size 1GByte.
         (FFT is applied to arrays of complex numbers, merging pairs of real H1 and L1 data.)
         Transforms $\tilde{Z}=(\tilde{H}_1,\tilde{L}_1)$ comprise $M$ sub-arrays ${\bf \tilde{Z}}_{k}$ $(k=1,2,\cdots, M)$, each of length $N$;
\item A chirp template $\bf w$ of duration $\tau=1$ s is extended by zeros to length $N$ and its transform $\bf \tilde{w}$ is loaded into {Global Memory}. 
         A {pre-callback function} computes $M$ transforms $\bf\tilde{\rho}$ from $M$ pointwise array multiplications ${\bf\tilde{\rho}}_{(k)} = {\bf\tilde{Z}}_{(k)}\cdot\bf \tilde{w}^*$ $(k=1,2,\cdots,M)$;
\item Inverse FFT$_{N,M}$ applied to ${\bf \tilde{Z}}_{(k)}$ produce $M$ corrections $\rho_k(t_n)$ over $N$ samples, representing the most computationally
        (but memory limited) intensive step on the GPU;
\item (ii) and (iii) are repeated $M$ times, once for each of $M$ chirp templates $\bf w$ at a total computational effort of $M^2$ inverse-FFT$_N$.
\end{enumerate}
At $5N\log_2N$ flops per FFT$_N$, these combined steps for above mentioned $N$ and $M^2$ involve $\sim$20 teraflop producing 2 TByte output. The latter shows the need to retain only tails of $M\times M$ convolutions ${\bf\tilde{\rho}}_{(k)}$ ($k=1,2,\cdots, M)$, i.e. candidate events exceeding a multiple of $\sigma_{km}$, one for each 16 s segment $k$ of data and chirp template $l$ $(k,m=1,2,\cdots,M)$. 

The $\sigma_{km}$ are pre-computed by Parseval's Theorem (\ref{EQN_PT}). As norms of complex Fourier coefficients, (\ref{EQN_PT}) is computationally demanding, requiring off-loading to the GPU as well (Fig. \ref{fig3}). For $\kappa=5.5$, for instance, tails are limited on the order of $10^4$ byte s$^{-1}$, well below the PCI bandwidth of several GByte s$^{-1}$, allowing near-optimal computing at about 65\% efficiency overall (including Parseval's step), normalised to FFT alone. Retaining tails over the PCI by the CPU is realized as follows.
 
\subsection{Gathering GPU-tails over the PCI}

\begin{figure}[h]
\centerline{\includegraphics[scale=0.45,angle=-90]{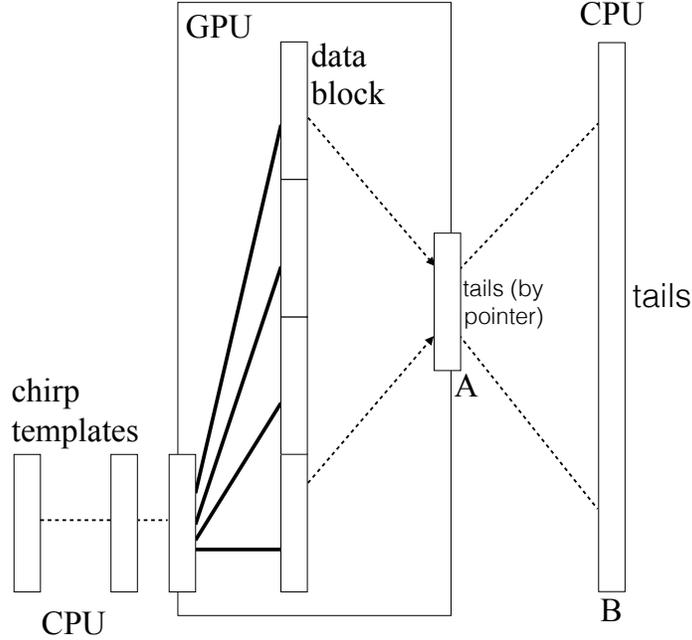}}
\caption{Tails of correlations between chirps with a block of $NM$ H1$\land$L1 (thick lines) projected by pointer into an array $A$ of size $N$ in Global Memory, that will sparse whenever $\kappa$ is on the order of a few. Gathered over the PCI, $*A$ are stored in an array $B$ of size $NM$ on the host. $B$ is stored to disk after completing correlations with a complete bank of chirp templates.}
\label{fig4}
\end{figure}

The tails of correlations satisfying (\ref{EQN_tail}) are gathered in two steps (Figs. \ref{fig3}-\ref{fig4}).
In correlations of a template ${\bf w}_k\, \epsilon\, W_\alpha$ and a segment ${\bf y}_{m}\,\epsilon Y_\beta$ ($1\le \alpha\le K_1$, $1\le\beta \le K_2$, $k,m=1,2,\cdots, M$), (\ref{EQN_tail}) is obtained for each $\sigma_{km}$. Thus, ${\bf w}_k$ gives $M$ tails in correlation with $Y_\beta$ referenced by time 
\begin{eqnarray}
t_{m} = t_{n_m+mN}
\label{EQN_tnm}
\end{eqnarray}
of maximal correlation satisfying
\begin{eqnarray}
\rho_{m}\ge\kappa\sigma_{km},
\label{EQN_C}
\end{eqnarray}
where $\rho_m = \rho(t_m)$.
To circumvent limited PCI bandwidth, (\ref{EQN_tnm}-\ref{EQN_C}) is converted to pointers projected into an array $A_{\alpha k}$ of size $N$,
\begin{eqnarray}
A_{\alpha k}=\{ (t_{m},\rho_{m}) | \, \rho_{m}\ge\rho_{m^\prime}\ge\kappa\sigma_{km^\prime} ~(\mbox{all}~m^\prime)\}.
\label{EQN_Ak}
\end{eqnarray}

We evaluate (\ref{EQN_Ak}) by post-callback function on the GPU by updating $(t_m,\rho_m)$ with $(t_{m^\prime},\rho_{m^\prime})$ whenever $\rho_{m^\prime}>\rho_m$ and $\rho_{m^\prime}>\sigma_{km^\prime}$. As an asynchronous read/write by pointwise index on Global Memory, this may lead to indeterministic behavior when two processors operate concurrently on the same index. When $\kappa$ is appreciable, $A_{\alpha k}$ is sparse, and this anomalous behavior is exceedingly rare. 

Repeating (\ref{EQN_Ak}) for all ${\bf w}_k\, \epsilon\, W_\alpha$ obtains $M^2$ tails by pointers
\begin{eqnarray} 
A_\alpha = \bigcup_{k=1}^M A_k.
\label{EQN_Aa}
\end{eqnarray}
Collecting all pointers in $A_\alpha$ is evaluated by the CPU. 

Gathering results over the complete template bank obtains by repeating (\ref{EQN_Aa}) for all $1\le \alpha \le K_1$, each time dereferencing $A_\alpha$ into an array $B$ of block size $NM$ on the host, 
\begin{eqnarray}
B = {\bigcup_{\alpha=1}^{K_1}} *A_\alpha,
\label{EQN_BA}
\end{eqnarray}
evaluated by the CPU. In collecting $B$, we select data with maximal $\rho$ values at $t_{n+mN}$ from the $*A_{\alpha}$.

Gathering {\em all} hits by removing selection of maximal $\rho$ in collecting $B$ in (\ref{EQN_BA}) produces extended output with up to two orders of magnitude more output in case of a signal. For a burst injection discussed below (Fig. 7), for instance, this increases output to tens of GByte for a bank of 8M templates. Such  extended output may be of interest to second runs, following up on selected data segments covering candidate events, but less so to first runs through all data such as LIGO S6.

\section{Benchmarks under OpenCL and filter output} \label{sec:3}

\begin{figure}[h]
\centerline{\includegraphics[scale=0.65]{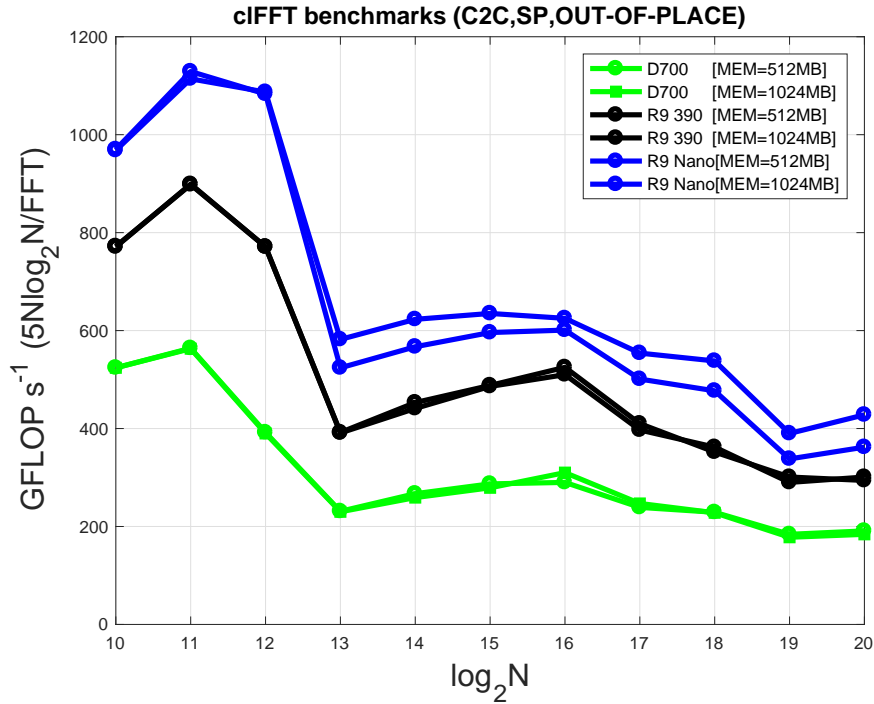}}
	\caption{Performance of FFT$_{N,M}$ by clFFT under OpenCL on the AMD GPU's D700, R9 390 and R9 Nano GPUs, 
	expressed in GFLOP\,s$^{-1}$ as a function of transform array size $N$ in C2C SP with 
	interleaved out-of-place data storage and no output back to the CPU. Results are shown for two different batch sizes
	with correspondingly different allocations in Global Memory. These results define a practical limit on performance in
	FFT-based correlations, that involve additional communications to a CPU over a PCI.}
	\label{fig5}
\end{figure}

The algorithm shown in Figs. \ref{fig3}-\ref{fig4} is implemented in Fortran90 and C++ using AMD's clFFT (in C99) under OpenCL. Following Table 2, clFFT operates on blocks of filtered H1$\land$L1 data in 1\,GByte blocks allocatable in Global Memory for clFFT$_{N,M}$ (C2C, SP) with interleaved out-of-place memory storage. 

Under OpenCL, a GPU is partitioned in CU's with fast but privy Local Memory and registers. Only {Global Memory} is shared across all CU's. Performance hereby critically depends on efficient use of Local Memory and minimal use of Global Memory, since access to the latter is relatively slow. With a Local Memory size of typically 32 kByte, clFFT performance for C2C SP will be essentially maximal  $N\le 2^{12}$. In our application, $N=2^{16}$, whereby clFFT performance is practically memory limited.

Fig. \ref{fig5} shows clFFT performance on GPUs with varying numbers of CUs (each comprising a number of {\em Stream Processors}) and Global Memory bus bandwidth (GByte s$^{-1}$), namely the R9 nano (4096, 64, 512), the R9 390 (2560, 40, 384) and the D700 (2048, 32, 264). For the first, performance is over 600 Gflop s$^{-1}$ for $N>2^{12}$ (about 1000 Gflop $s^{-1}$ for $N\le 2^{12}$). This is a direct result of the 32 kByte Local Memory size and $8N$ bytes in complex single precision storage and the need to access Global Memory when $N>2^{12}$. For $N=2^{16}$, the net result is overall efficiency of about 7\% of peak floating point compute-performance by Stream Processors alone. 

We implement Parseval's Theorem by partial sums off-loaded to the GPU, the results of which are summed by the CPU. At a few hundred Gflop\,s$^{-1}$ performance thus achieved, wall clock compute time is about 25\% compared to that of clFFT on the GPU. Including overhead in ($i-iv$) of \S3.2 and gathering tails (\S3.3), the net result (including Parseval's step) is an efficiency overall of about 65\%, normalised to clFFT alone as shown in Fig. \ref{fig4}, or about $\sim 8\times 10^4$ correlations per second per GPU. On a cluster of about a dozen GPU's, we hereby realise about 1 million correlations per second, sufficient for a real-time analysis by up to 16 million templates according to (\ref{EQN_nd}).

Filter output stored to disk is listed by block in files B$n$, $n=1,2,\cdots, 288$, illustrated in Table 3.

\begin{table}
{\bf Table 3.} {Butterfly filtering output B$n$ of a block $n$ $(n=1,2,\cdots,288)$ of hits $\rho_i>\kappa\sigma$ $(i=1,2)$
lists data sample offset $i\,\epsilon\,\{1,2,\cdots, 2^{27}\}$, $\rho_i$ and $f_i$, the latter the initial frequency of
associated chirp template. Multiplication of $\rho_i$ by 1000 allows storage of all entries in 4 byte integers. Sample
shown of B161 (6,388,647 rows produced by a bank of 4M templates)
highlights some simultaneous hits. Zeros represent no hit.}
\center{	{\begin{tabular}{@{}cccccc@{}}
\mbox{}\\\hline\hline
Sample offset $i$ & $1000\times \rho_i$(H1) & $1000\times \rho_i$(L1) & $f_i$(H1) [Hz] & $f_i$(L1) [Hz] \\
\hline
$\cdots$ \\
17712959 & 0 & 5522 & 0 & 1988 \\ 
17713193 & 5747 & 0 & 486 & 0 \\ 
17713194 & 5516 & 0 & 623 & 0 \\ 
17713195 & 6424 & 0 & 632 & 0 \\ 
17713196 & 6578 & 6660 & 497 & 489 \\ 
17713197 & 5769 & 7491 & 488 & 489 \\ 
17713198 & 7315 & 6671 & 490 & 489 \\ 
17713199 & 8530 & 7111 & 563 & 565 \\ 
$\cdots$ \\
\hline
\end{tabular}}}
\label{Table3}
\end{table}

\begin{figure}[h]
\centerline{\includegraphics[scale=0.48]{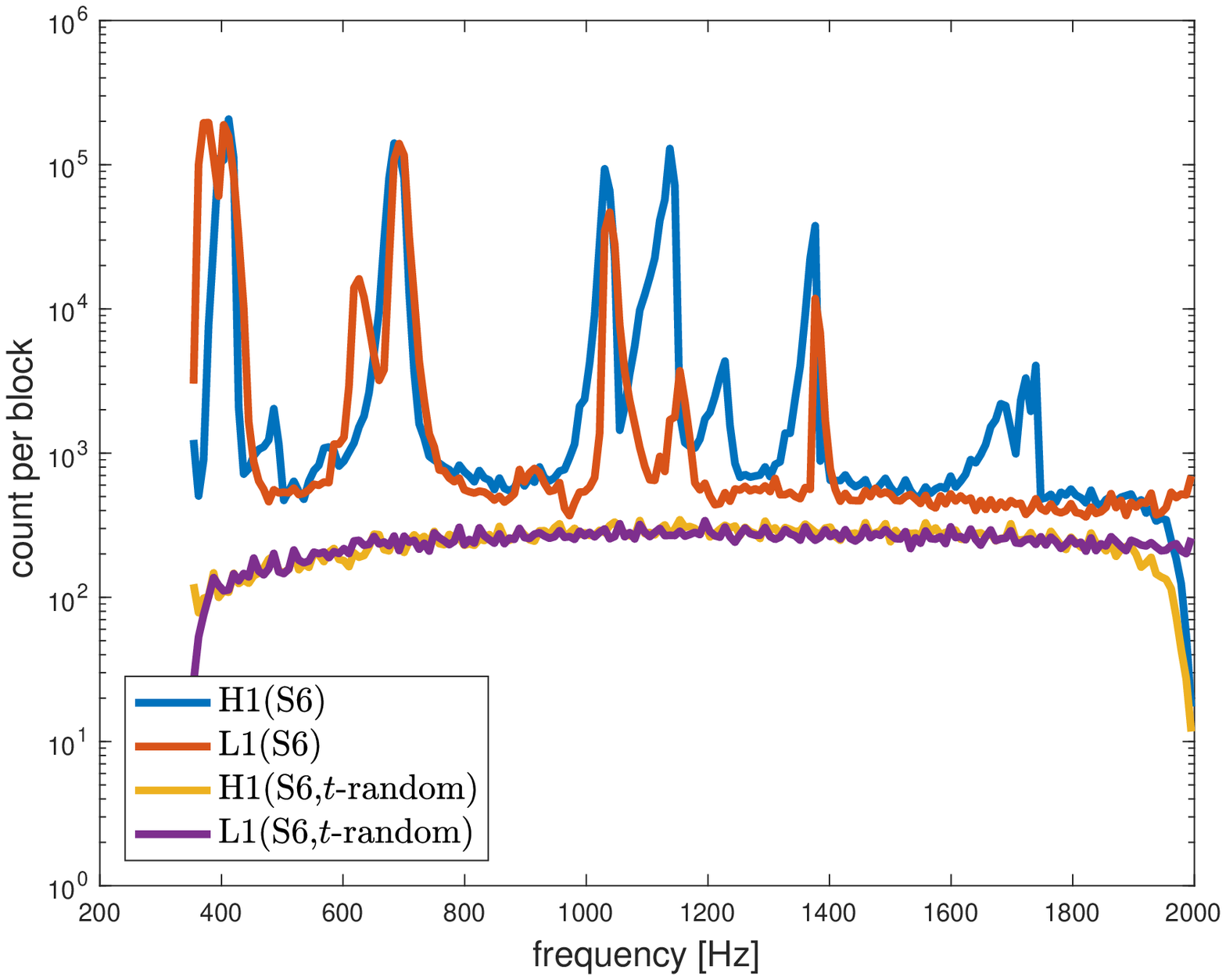}\hfill\includegraphics[scale=0.48]{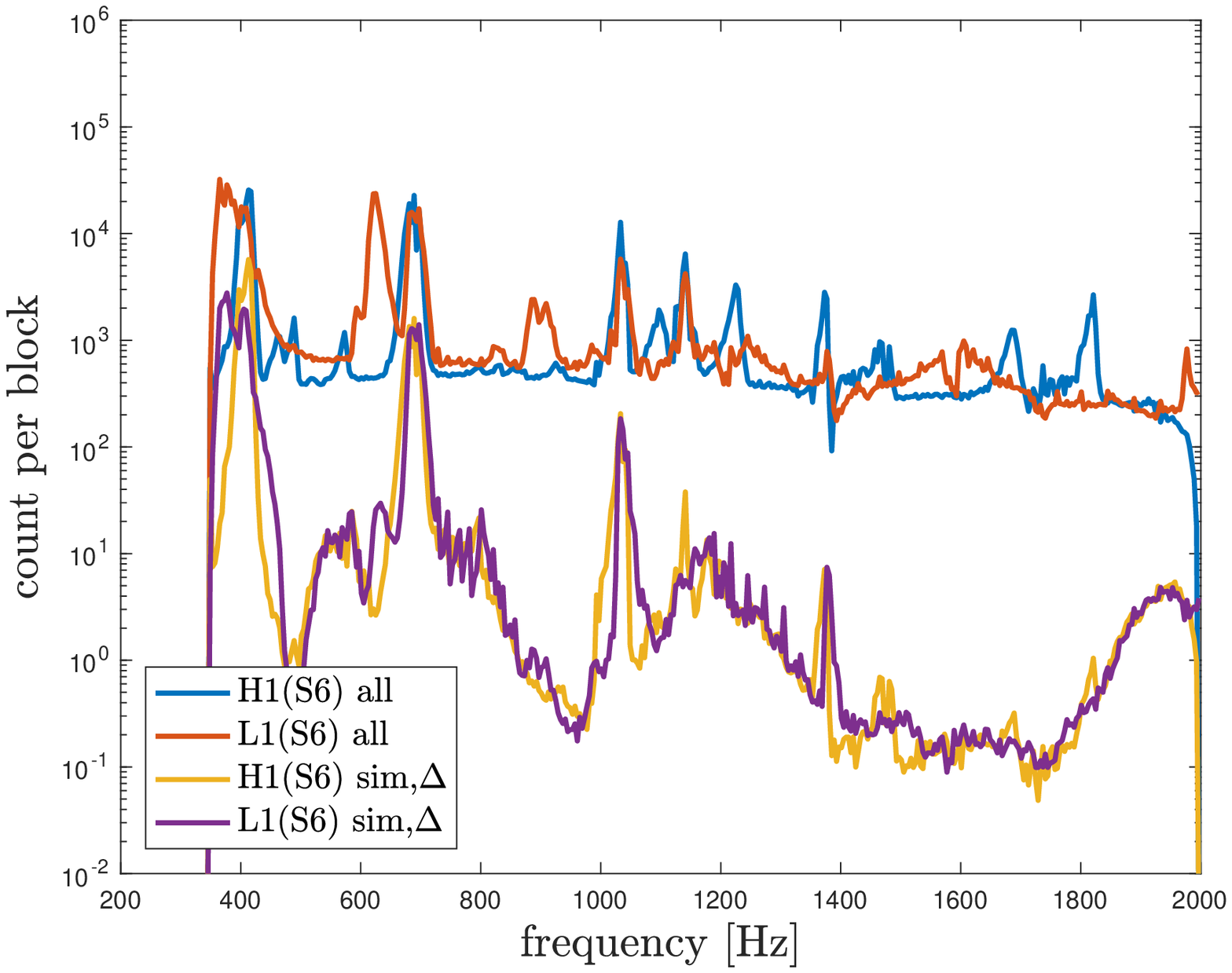}}
\caption{(Left panel.) Pseudo-spectra of simultaneous hits in tails $>\kappa\sigma$ with $\kappa=5.5$ of butterfly filtered output of H1 (red) and L1 (blue), shown as an average over four blocks (161-2,177-8) of S6 H1$\land$L1, using a bank of type A of 8M chirp templates, along with baseline results following time-randomized data. (Right panel.) Pseudo-spectra as an average over {\em all} 288 blocks of S6 H1$\land$L1, using a bank of type A of 0.5M chirp templates, of H1 and L1 by independent counts and by simultaneous counts with frequency pairs within $\Delta = 50$\,Hz.}
\label{fig6}
\end{figure}
\begin{figure}[h]
\centerline{ \includegraphics[scale=0.405]{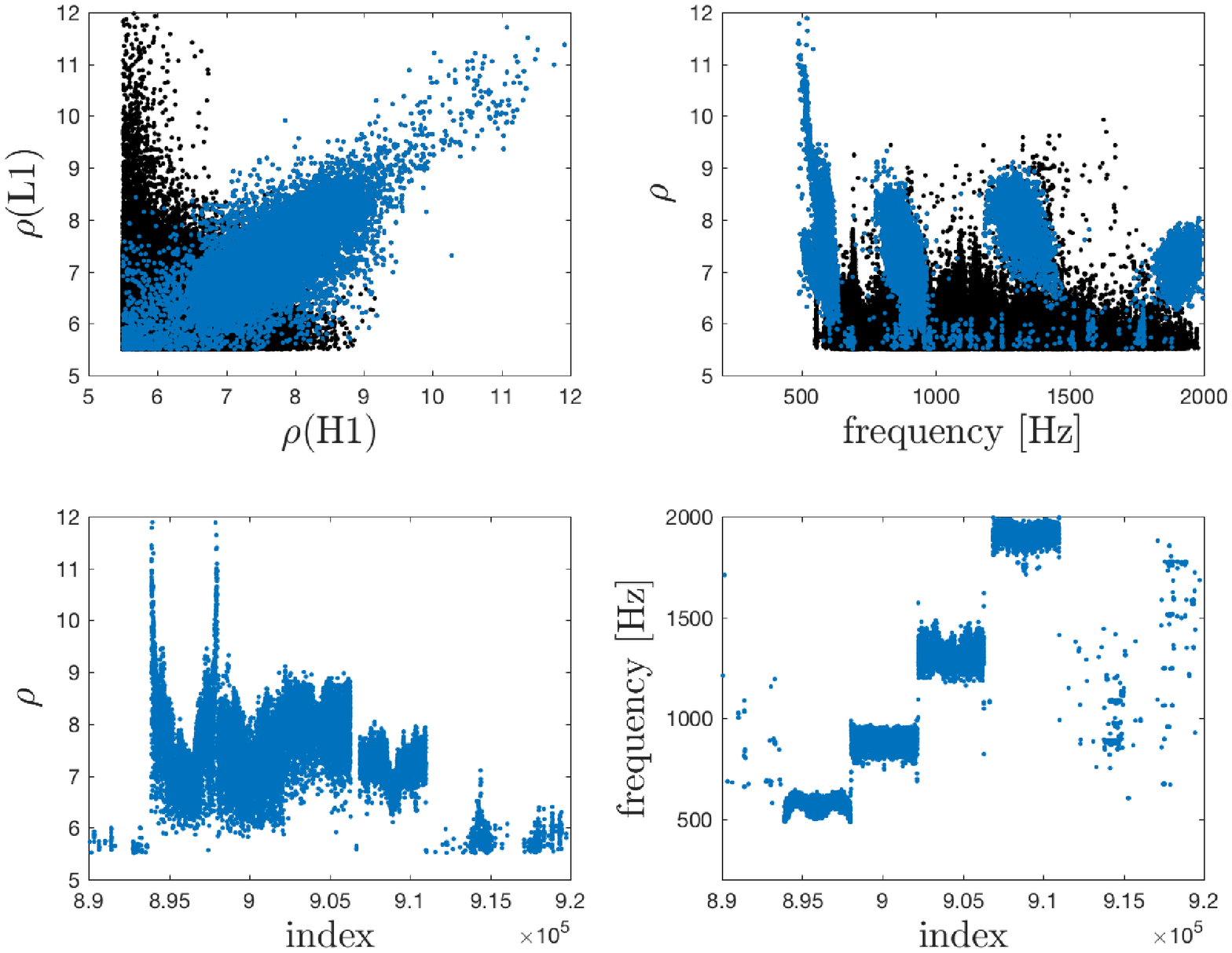}\includegraphics[scale=0.42]{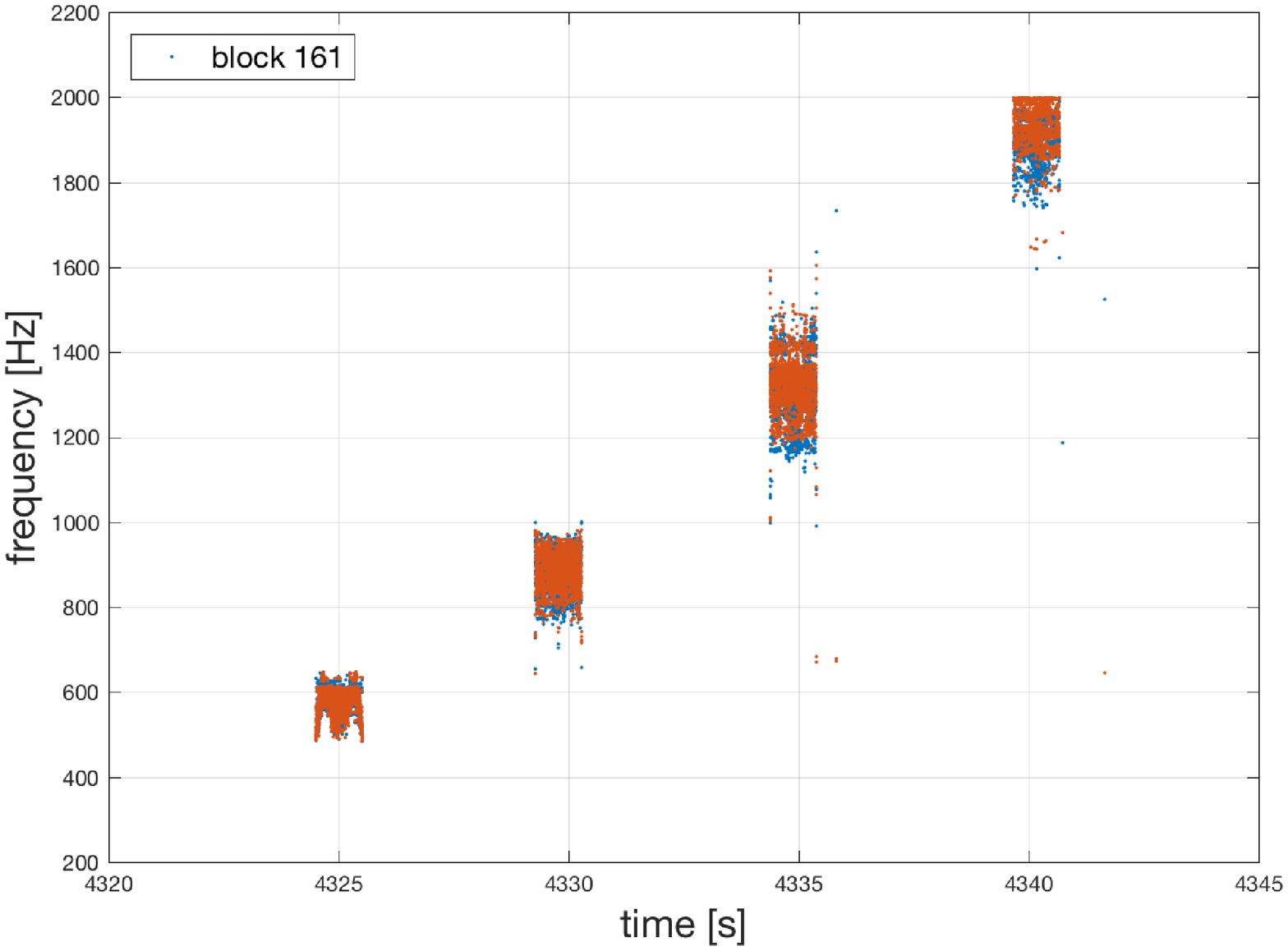}}
\caption{Detection of a high-frequency LIGO injection in block 161 by butterfly filtering with bank of type B of 4M chirp templates at large injection SNR (Table 4), seen in simultaneous hits in H1 and L1 (left panels; $\rho$ and frequency refer to geometric means of those of H1 and L1). Hits in H1 (red) and L1 (blue) practically overlap (right panel) shown as a function of time based on data sample offset in output file B161 (Table 3).}
\label{fig7}
\end{figure}

\begin{table}
{\bf Table 4.} {Sample of a LIGO S6 injection in H1 and L1 \citep{losA}, comprising a sequence of sine-Gaussian signals \citep{mot10} stepwise covering 50-2000 Hz with injection signal-to-noise ratio SNR listed by the LOSC.}
\center{	{\begin{tabular}{@{}lclc@{}}
\mbox{}\\\hline\hline
 GPS time [s] & strain amplitude ($h_{rss}$) & Waveform ($f$[Hz], $Q$) & SNR(LOSC)  \\
\hline
H1 \\
\hline
958413408.20 & $3.57\times10^{-21}$ &   sine-Gaussian (393,9)  & 93.27 \\
958413413.50 & $4.33\times10^{-21}$ &   sine-Gaussian (554,9)   & 92.37\\
958413418.30 & $6.41\times10^{-21}$ &   sine-Gaussian (850,9)   & 98.23\\
958413423.40 & $9.84\times10^{-21}$ &   sine-Gaussian (1304,9)  & 91.20\\
958413428.70 & $7.47\times10^{-21}$ &   sine-Gaussian (2000,9)  & 23.15\\
\hline
L1 \\ \hline
958413408.20 & $3.63\times10^{-21}$ &   sine-Gaussian (393,9)    & 65.37\\
958413413.50 & $4.74\times10^{-21}$ &   sine-Gaussian (554,9)    & 68.50\\
958413418.30 & $7.19\times10^{-21}$ &   sine-Gaussian (850,9)   & 72.22\\
958413423.40 & $1.09\times10^{-20}$ &   sine-Gaussian (1304,9)  & 67.98\\
958413428.70 & $8.61\times10^{-21}$ &   sine-Gaussian (2000,9)  & 25.85\\
\hline
\end{tabular}}}
\label{Table4}
\end{table}

\begin{figure}[h]
\center{\includegraphics[scale=0.65,angle=90]{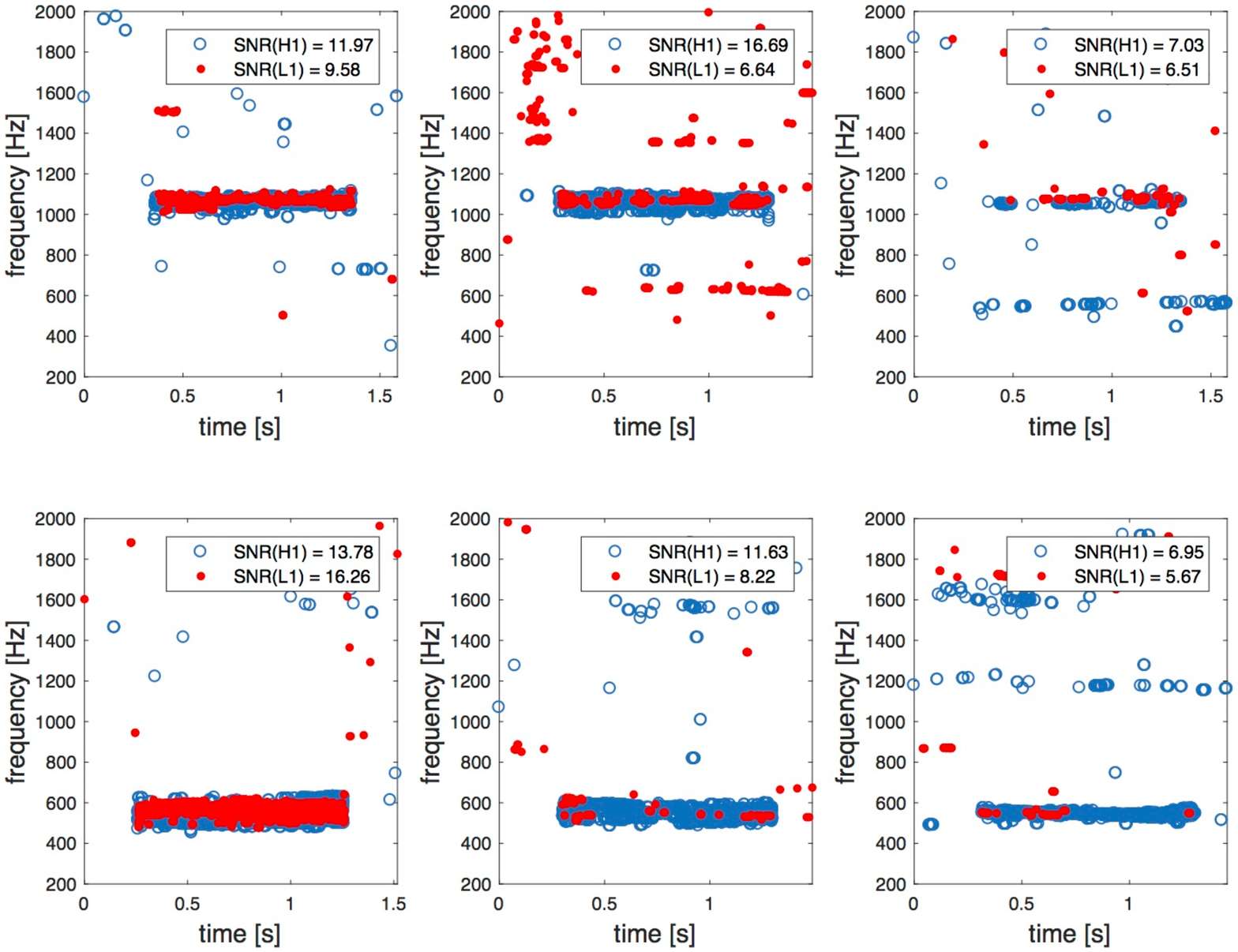} 
\caption{Selected high frequency injections at 1053\,Hz (top panels) and 554\,Hz (bottom panels) H1$\land$L1 LIGO S6 at high to low injection SNR(LOSC) (left-to-right columns), here detected by H1 (blue circles) and L1 (red dots) using a bank of type A of 8M chirp templates. The 1053 Hz (554 Hz) injections shown are at the respective GPS times 932380188.50, 935143367.60, 946193522.10 (959322411.40,  934962011.00, 946205393.30). }}
\label{fig8}
\end{figure}
\begin{figure}[h]
\center{\includegraphics[scale=0.65]{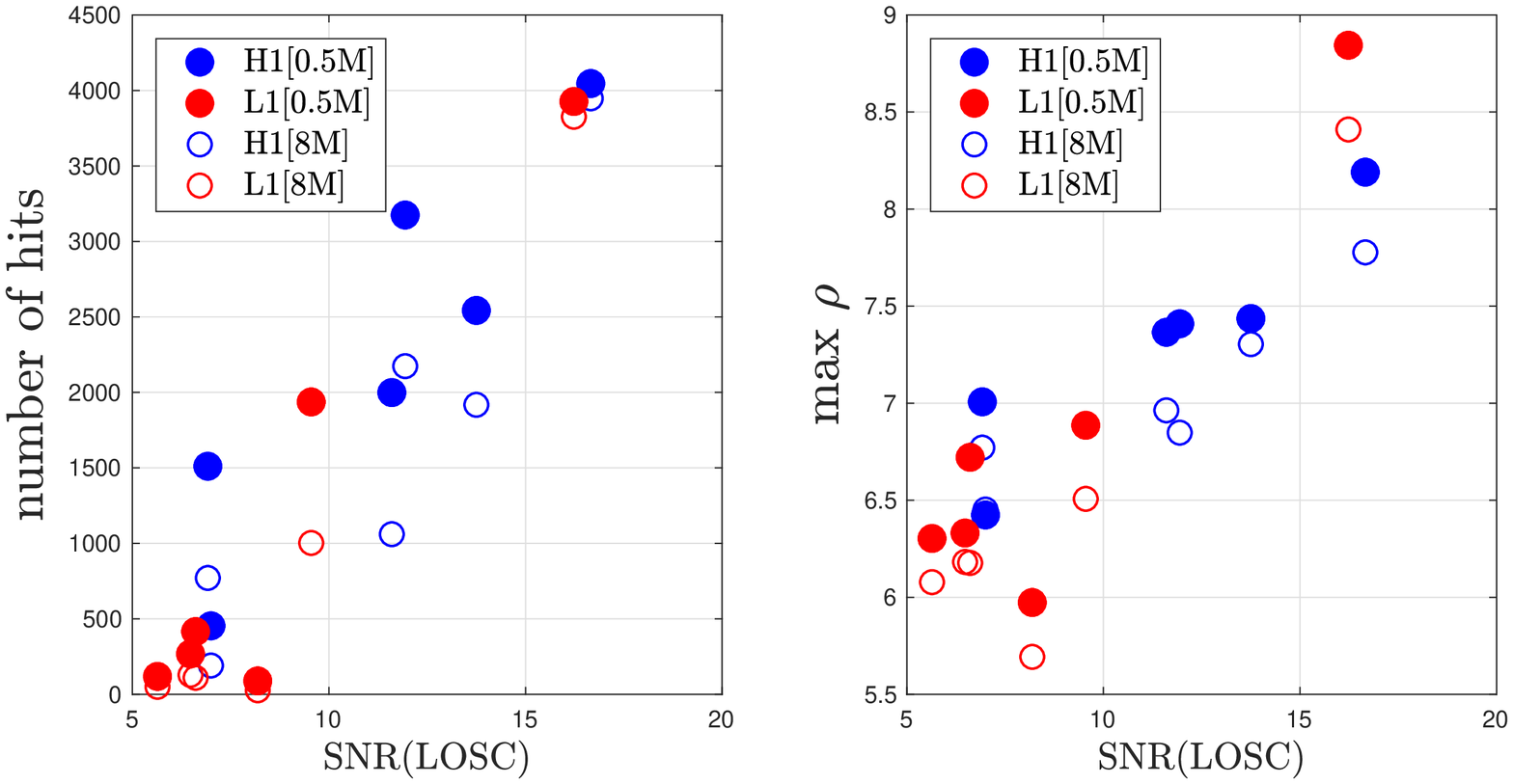} 
\caption{The number of hits $\rho(t)>\kappa \sigma$ and maximal values of $\rho$ in Fig. 8 about injections at 554 Hz and 1053 Hz in H1 and L1 shows a generic trend with SNR in the injection process. On average, counts improve by a factor of 1.69 and $\rho$ increments by 0.29 with the template bank of 8M compared to 0.5M chirps. Hits are counted with a frequency margin of $\pm 50$\,Hz about these injection frequencies.}}
\label{fig9}
\end{figure}

\section{Tails and LIGO burst injections}

To illustrate a full analysis, Fig. \ref{fig6} shows a pseudo-spectrum of the tails (\ref{EQN_tail}) of H1$\land$L1 LIGO S6, obtained
by averaging results of blocks using a template bank of intermediate size of 0.5\,M chirps. The detailed structure shown represents the {\em non}-Gaussian features that carry any potentially relevant information, visible only by zooming in on tails in an otherwise overall near-Gaussian PDF of the internal GPU output $\rho$ (Fig. \ref{fig3}). This has been verified numerically, in obtaining completely smooth spectra of tails of $\rho$ following time-randomisation of H1 or L1 data (Fig. \ref{fig6}).

Fig. \ref{fig6} shows various pronounced features, some of which are probably associated with unsteady behaviour in various instrumental lines familiar from conventional Fourier spectra of S6 strain noise \citep{losS}. The details of which remain to be understood in more detail,
especially so given the non-trivial residual spectrum of simultaneous hits with frequency pairs $(f_1,f_2)$ of H1 and L1 that are relatively close, here shown with $\left| f_1 -f_2 \right| < \Delta$, $\Delta = 50$\,Hz. (A similar spectrum obtains for $\Delta = 100\,$ Hz.) For the analysis with a bank of 0.5M templates shown, the total counts per block for H1 and L1 are $(2.1 \times 10^6,3.2\times 10^6)$ with simultaneous counts (19.73\%,13.01\%), reduced to (6.82\%,4.49\%) for frequency pairs within $\Delta = 50$\,Hz ((7.55\%,4.98\%) for frequency pairs within $\Delta = 100$\,Hz).

LIGO detectors are routinely given a variety of {hardware injections} to test the detectors and various signal detection pipelines. Of interest to the present analysis are burst injections that cover the relatively high frequency range 350-2000 Hz. The following uses some LIGO injections for a formal test and validation of software implementation.

Fig. \ref{fig7} shows an injection to both H1 and L1 captured by our algorithm at large injection SNR(LOSC), detected using a bank of 4M templates in a partial analysis of LIGO S6.

For high-confidence detections, correlated H1-L1 output such as illustrated in Fig. \ref{fig7} is essential. While signal injections are often injected at the same GPS time, astrophysical sources will impact H1 and L1 along some finite viewing angle. In the time-domain, this is commonly identified by maximising correlations over a some finite time shift, here 0-10 ms given the distance between H1 and L1. Here, we make use of the fact that a difference in arrival time between H1 and L1 from a putative astrophysical source with finite time rate-of-change in $f(t)$ is equivalent to a frequency shift, allowing searches in simultaneous H1-L1 filter output such as plotted in Fig. \ref{fig7}.

Figs. \ref{fig8}-\ref{fig9} shows a validation of sensitivity (see also earlier analyses of \cite{van14a,van16}), here a priori limited to $\rho$ exceeding 5.5\,$\sigma$ by choice of $\kappa$ in (\ref{EQN_tail}), obtained in a partial LIGO S6 analysis using 8M templates. Overall, it appears that sensitivity in H1 is slightly better than L1 when signals are small.  Searches for signals fainter than those shown would require a re-run of the analysis with $\kappa<5.5$ in (\ref{EQN_tail}). For such extremely deep searches, excess tail sizes can conceivably be curtailed by generalising (\ref{EQN_tail}) to a finite band, $\kappa_1 \sigma > \rho(t_n) > \kappa_2 \sigma$ with $\kappa_2-\kappa_1\lesssim1$.

Fig. \ref{fig9} quantifies the gain in using bank sizes beyond the minimal requirements (\S3.1), showing an increase in hit counts and $\rho$ in a detection of a sample of high frequency burst injections. 

\section{Conclusions and outlook}
\label{sec:6}

Probing inner engines to gamma-ray bursts and core-collapse supernovae require deep searches in LIGO data. Taking full advantage of modern GPU hardware, we present a GPU-CPU implementation of butterfly filtering to search for broadband extended emission in gravitational waves from accreting flows around black holes, potentially relevant to the most extreme transient events.

Our benchmarks demonstrate near-optimal performance using banks of up to millions of chirp templates at better than real-time analysis, facilitating deep searches in LIGO archive data such as S6, advanced LIGO O1 and the currently ongoing O2 run. 

Specific applications of the proposed method include correlation analysis of the H1 and L1 detectors and identification of mysterious or peculiar events of interest to further analysis. A leading order indication of correlations may derive, for instance, from counting statistics of hits, comparing simultaneous hit counts with total hit counts in H1 and L1. Specific events of interest may be followed up by second runs, gathering all hits by removing selection of maxima in collecting $B$ in (\ref{EQN_BA}). 

In butterfly filtering, signal detection typically comprises a large number hits, representing approximate matches with no single template providing a perfect match to the full signal at hand, as illustrated in Fig. 8. This combined output can in principle recover essentially maximal sensitivity
\citep{van16}. For automated searches of candidate events, clustering algorithms might apply \citep[e.g][]{geo17}, that may also facilitate quantifying the level of confidence for such complex detection output.

{\bf Acknowledgments.} The author gratefully acknowledges detailed constructive comments from the referee and J.B. Kanner. This work was partially supported by the National Research Foundation of Korea under grants 2015R1D1A1A01059793 and 2016R1A5A1013277 and made use of LIGO S6 data from the LIGO Open Science Center (losc.ligo. org), provided by the LIGO Laboratory and LIGO Scientific Collaboration. LIGO is funded by the U.S. National Science Foundation. Additional support is acknowledged from MEXT, JSPS Leading- edge Research Infrastructure Program, JSPS Grant-in- Aid for Specially Promoted Research 26000005, MEXT Grant-in-Aid for Scientific Research on Innovative Areas 24103005, JSPS Core-to-Core Program, A. Advanced Re- search Networks, and the joint research program of the Institute for Cosmic Ray Research.

\end{document}